\documentclass{JHEP3}
\usepackage{epsfig}

\def\lsi{\raise0.3ex\hbox{$<$\kern-0.75em\raise-1.1ex\hbox{$\sim$}}}
\def\gsi{\raise0.3ex\hbox{$>$\kern-0.75em\raise-1.1ex\hbox{$\sim$}}}

\title{Lattice determination of the critical point of 
QCD at finite $T$ and $\mu$}
\author{Z.~Fodor\\
Deutsches Elektronen-Synchrotron DESY, Notkestr. 85, D-22607, 
Hamburg, Germany\\
Institute for Theoretical Physics, E\"otv\"os University, P\'azm\'any
1, H-1117 Budapest, Hungary.\\ Email:\email{fodor@poe.elte.hu}}
\author{S.D.~Katz\\
Institute for Theoretical Physics, E\"otv\"os University, P\'azm\'any
1, H-1117 Budapest, Hungary.\\ Email:\email{katz@bodri.elte.hu}}

\abstract{ 
Based on universal arguments it is believed that there is a 
critical point (E) in QCD on the temperature ($T$) versus chemical
potential ($\mu$) plane, which is of extreme importance for
heavy-ion experiments. Using finite size scaling and 
a recently proposed lattice method to study QCD at finite $\mu$
we determine the location of E
in QCD with $n_f$=2+1 dynamical staggered quarks 
with semi-realistic masses
on $L_t=4$ lattices. Our result is $T_E=160 \pm 3.5$~MeV and 
$\mu_E= 725 \pm 35$~MeV. For the critical temperature at 
$\mu=0$ we obtained $T_c=172 \pm 3$ MeV.}


\preprint{ITP-Budapest 568\\ DESY 01-057}

\begin{document}
\section{Introduction.}
 
QCD at finite $T$ and/or $\mu$ is of fundamental importance,
since it describes relevant features of particle physics
in the early universe, in neutron stars and in heavy ion collisions 
(for a clear introduction see \cite{W00}).
QCD is asymptotically free, thus its high $T$ and high density
phases are dominated by partons (quarks and gluons) as degrees of
freedom rather than hadrons. In this quark-gluon plasma (QGP) phase
the symmetries of QCD are restored. In addition, 
recently a particularly interesting, rich phase structure has been 
conjectured for QCD at finite $T$ and $\mu$ 
\cite{qcd_phase,crit_point}. 

Extensive experimental work has been done
with heavy ion collisions at CERN and Brookhaven to explore
the $\mu$-$T$ phase diagram. Note, that 
past, present and future heavy ion 
experiments with always higher and higher energies produce states
closer and closer to the $T$ axis of the $\mu$-$T$ diagram. It is
a long-standing open question, whether a critical point
exists on the $\mu$-$T$ plane, 
and particularly how to predict theoretically its location 
\cite{crit_point,SKP99}. 

Let us discuss first the $\mu$=0 case. 
Universal arguments \cite{PW84} and lattice simulations \cite{U97} 
indicate that in a hypothetical QCD
with a strange (s) quark mass ($m_s$) as small as the up (u) and down (d) 
quark masses ($m_{u,d}$) 
there would be a first order finite 
$T$ phase transition. The other extreme case ($n_f$=2)
with light u/d quarks but with an infinitely large $m_s$ 
there would be no phase transition only an analytical
crossover. Note, that observables change rapidly during a crossover,
but no singularities appear (we will use the expression ``transition'' 
if we do not want to specify whether we deal with a phase transition
or a crossover). This means that between the two extremes there is a 
critical strange mass ($m_s^c$) at which one has a second order finite 
$T$ phase transition. Staggered lattice results on $L_t$=4 lattices  
with two light quarks and $m_s$ around the transition $T$ ($n_f$=2+1)  
indicated \cite{B90} that $m_s^c$ is about half of the physical $m_s$. 
Thus, in the real world we probably have a crossover. (Clearly,
more work is needed to approach the chiral and continuum limits. Note,
that the puzzle due to an unexpected strengthening 
observed \cite{B94} for the $n_f$=2 
Wilson action with intermediate $m_{u,d}$ 
was resolved by using an improved action \cite{I97}.) 
           
Returning to a non-vanishing $\mu$, one realizes that arguments 
based on a variety of models (see e.g. \cite{B89,qcd_phase,crit_point})
predict a first order finite $T$ phase transition at large $\mu$. 
Combining the $\mu=0$ and large $\mu$ informations an interesting
picture emerges on the $\mu$-$T$ plane. For the physical $m_s$
the first order phase transitions at large $\mu$ should be connected
with the crossover on the $\mu=0$ axis. This suggests
that the phase diagram features a critical endpoint $E$ (with
chemical potential $\mu_E$ and temperature $T_E$), at which  
the line of first order phase transitions ($\mu>\mu_E$ and $T<T_E$) 
ends \cite{crit_point}. At this point the phase transition is of
second order and long wavelength fluctuations appear, which 
results in characteristic experimental consequences, similar to
critical opalescence. Passing close enough to ($\mu_E$,$T_E$)  
one expects simultaneous appearance of 
signatures (e.g. freeze-out type behavior of
observables constructed from the
multiplicity and transverse momenta of charged pions),
which exhibit nonmonotonic dependence on the 
control parameters \cite{SRS99}, 
since one can miss the critical point on either of two sides.
 
The location of this critical point is 
an unambiguous, non-perturbative prediction of the QCD Lagrangian. 
Unfortunately, no
{\it ab initio}, lattice analysis based on QCD was done to locate
the endpoint. Crude models 
with $m_s=\infty$ were used (e.g. \cite{crit_point})
suggesting that $\mu_E \approx$ 700~MeV, which should be smaller
for finite $m_s$.  The result is sensitive to $m_s$,
thus for realistic cases previous 
works could not predict the value of $\mu_E$ 
even to within a factor of 2-3.  The goal of this
exploratory work is to show how to locate the endpoint by a lattice
QCD calculation. We use full QCD with dynamical $n_f$=2+1 
staggered quarks.      

QCD at finite $\mu$ can be
formulated on the lattice \cite{HK83}; however, standard
Monte-Carlo techniques can not be used at $\mu \neq 0$. The reason
is that for non-vanishing real $\mu$ the functional measure
--thus, the determinant of
the Euclidean Dirac operator-- is complex. This fact
spoils any Monte-Carlo technique based on importance sampling.
Several suggestions were studied to solve the problem.
Unfortunately, none of them was able to locate ($\mu_E$,$T_E$).

In a recent paper we proposed a new 
method \cite{FK01} to study
lattice QCD at finite $T$ and $\mu$. The 
idea was to produce an ensemble of QCD configurations at
$\mu$=0 and at $T_c$. Then we determined
the Boltzmann weights \cite{FS89} of these configurations at $\mu\neq 0$
and at $T$ lowered to the transition temperatures at this
non-vanishing $\mu$. Since transition configurations 
were reweighted to transition configurations a much better
overlap was observed than by reweighting pure hadronic configurations     
to transition ones \cite{glasgow}. We illustrated the applicability 
of the method in $n_f$=4 dynamical QCD. Using  
only ${\cal {O}}(10^3$-$10^4)$ configurations quite
large $\mu$ could be reached and the transition line
separating the hadronic phase and the QGP was given
on the $\mu$-$T$ plane.

In this letter we generalize the above method to arbitrary number of
staggered quarks. We apply it to the $n_f$=2+1 case. We determine
the volume (V) dependence of the Lee-Yang zeros of the partition function 
on the complex gauge coupling ($\beta$) plane. Based on this V
dependence we determine the type of the transition as a function of
$\mu$. The endpoint $\mu_E$ is given by the
value at which the crossover disappears and finite-V
scaling predicts a first order phase transition. These finite
$T$ calculations are done on $L_t=4$ lattices. In order 
to set the physical scale we determine the pion and rho masses
($m_\pi,m_\rho$), the string-tension ($\sigma$) and the
Sommer \cite{S94} scale ($R_0$) at $T=0$. 
Our quark masses are ``semi-realistic'':
$m_s$ is set about to its physical value, whereas 
$m_{u,d}$ are approximately four times as heavy as they are in  
the real world. Having determined the lattice spacing we transform
our result to physical units and give $T_c$
the location of ($\mu_E$,$T_E$) and show the phase diagram separating
the hadronic phase and the QGP.

Though this study performes a V$\rightarrow \infty$ extrapolation, 
larger volumes, larger $L_t$-s
and smaller masses are also needed to give the final answer to
($\mu_E$,$T_E$).

\section{Staggered quarks at $\mu\neq 0$. }

The partition function of QCD with $n_f$ degenerate 
staggered quarks (for an introduction see eg. \cite{MM94}) is
given by the functional integral of the bosonic action
$S_{b}$ at gauge coupling $\beta$ over the link variables $U$, 
weighted by the determinant of the quark matrix $M$, which can be 
rewritten \cite{FK01} as
\begin{eqnarray} \label{reweight}
&&Z(\beta,m,\mu)=
\int{\cal D}U \exp[-S_{b}(\beta,U)][\det M(m,\mu,U)]^{n_f/4}
\nonumber\\
&&=\int {\cal D}U \exp[-S_{b}(\beta_w,U)][\det M(m,\mu_w,U)]^{n_f/4}
\\
&&\left\{\exp[-S_{b}(\beta,U)+S_{b}(\beta_w,U)]
{[\det M(m,\mu,U)]^{n_f/4}  \over [\det M(m,\mu_w,U)]^{n_f/4}}\right\},
\nonumber
\end{eqnarray}
where $m$ is the quark mass, $\mu$ is the chemical potential of the quark. 
For non-degenerate masses one uses simply the product of several quark matrix
determinants on the $1/4$-th power. Standard
importance sampling works and can be used to collect an 
ensemble of configurations at  $\beta_w$ and $\mu_w$ with  
Re($\mu_w$)=0. It means we treat the terms in the curly 
bracket as an observable
--which is measured on each independent configuration--
and the rest as the measure. By simultaneously changing
$\beta$ and $\mu$ one can ensure that even the mismatched measure
at $\beta_w$ and $\mu_w$ samples 
the regions where the original integrand with $\beta$ and $\mu$ 
is large. In practice the determinant is evaluated at some $\mu$ and 
a Ferrenberg-Swendsen reweighting \cite{FS89} is performed
for the gauge coupling $\beta$. 

Due to the complex nature of $\det M(m,\mu,U)$ 
an additional problem arises, one should 
choose among the possible Riemann-leaves of the fractional power
in eq. (\ref{reweight}). This can be done by using 
the fact that at $\mu=\mu_w$ the ratio of the determinants is 1 and
the ratio should be a continuous function of $\mu$.
However, the continuity can only be ensured if the analytical
dependence of the determinant on $\mu$ is known (the determinant oscillates
strongly with $\mu$, so measuring it for several $\mu$ values is not 
satisfactory). This dependence can be given by the following way (the 
idea goes back to a method of \cite{Toussaint}).

First gauge fix to $A_0=0$ on all but the last timeslice
\begin{equation}
M=\left(
\begin{array}{ccccc}
B_0         & e^{\mu  }  & 0         & \dots & e^{-\mu  } T^{\dagger} \\
-e^{-\mu  } & B_1        & e^{\mu  } & \dots & \\
0	    &-e^{-\mu  } & B_2       & \dots & \\
\vdots      & \vdots     & \vdots    & \ddots& \\
-e^{\mu  } T & & & & \\
\end{array}
\right).
\end{equation}
The $B_i$ are $3L_s^3 \times 3L_s^3$ matrices containing the mass and spatial
hopping terms (3 is the number of colors)
and $T$ contains the only remaining temporal links on 
the last timeslice. By multiplying the $j$-th column of M by $e^{-j\mu }$ and
the $j$-th row by $e^{j\mu }$ and moving the leftmost column to the right
one gets a matrix with the same determinant
\begin{equation}
\left(
\begin{array}{ccccc}
1  & 0         & \dots & e^{-L_t \mu } T^{\dagger} & B_0         \\
B_1        & 1 & \dots & 0 & -1\\
\vdots      & \vdots   & \ddots& &\vdots\\
& & & 1 &0 \\
& & & B_{L_t-1}& -e^{L_t \mu } T \\
\end{array}
\right)
\end{equation}
We evaluate the determinant by 
Gauss-elimination. After $L_t-2$ steps we get 
a $6 L_s^3 \times 6 L_s^3$ matrix
\begin{equation}
\left(
\begin{array}{cc}
1+c_1\cdot e^{-L_t\mu} & \;\;\; c_2 \\
c_3+c_4\cdot e^{-L_t\mu}  & \;\;\; c_5- e^{L_t\mu}T \\
\end{array}
\right),
\end{equation}
where $c_i$ are 
$\mu$-independent matrices. It is straightforward to show that
\begin{equation}
\det M=e^{-3 L_s^3 L_t\mu} \det (P-e^{L_t\mu}), 
\end{equation}
where 
\begin{equation}
P= \left(
\begin{array}{cc}
-c_1 & \;\;\; c_2 T^{\dagger} \\
c_3 c_1- c_4 &\;\;\; (c_5 - c_3 c_2) T^\dagger \\
\end{array}
\right).
\end{equation}
This is just the characteristic equation of $P$. 
To find the $\lambda_i$ eigenvalues one needs 
${\cal O}(L_s^9)$ 
operations, which gives the determinant as a function of $\mu$ and
determines the ratio of the fractional powers 
in eq. (\ref{reweight}) unambiguously
\begin{equation}\label{det}
\det M(\mu)=e^{-3 L_s^3 L_t \mu } \prod_{i=1}^{6 L_s^3} (e^{L_t \mu }-\lambda_i).
\end{equation}

Using eq. (\ref{reweight}) with the determinant given by
eq. (\ref{det}) we can give the partition function for 
complex $\mu$ and $\beta$ values.  In the 
following we keep $\mu$ real and look for the zeros of the partition
function on the complex $\beta$ plane. These are the 
Lee-Yang zeros \cite{LY52}, whose V$\rightarrow \infty$
behavior tells the difference between a crossover and a first order
phase transition. At a first order phase transition the free energy
$\propto \log Z(\beta)$ is non-analytic.
Clearly, a 
phase transition can appear only in the V$\rightarrow \infty$ limit, 
but not in a finite $V$. Nevertheless, the partition
function has zeros at finite V, the Lee-Yang zeros, 
at ``unphysical'' complex values
of the parameters, in our case at complex $\beta$. For a 
system with a first order phase transition these zeros
approach the real axis in the V$\rightarrow \infty$ limit
(detailed analyzes suggest $1/V$ scaling).   
This V$\rightarrow \infty$ limit generates the non-analyticity of
the free energy. For a system with crossover the free energy is analytic, 
thus the zeros do
not approach the real axis in the V$\rightarrow \infty$ limit.
The Lee-Yang technique was successfully applied to determine the 
endpoint of the electroweak phase transition \cite{3ewpt,4ewpt}.

\section{$T \neq 0$ and $T=0$ results for $n_f$=2+1.}

Using the formulation described above we study $n_f$=2+1
QCD at $T\neq 0$ on $L_t=4,\ L_s=4,6,8$ lattices and at $T=0$
on $V=10^3\cdot 16$ lattices. $m_{u,d}=0.025$ and $m_s=0.2$ were
chosen for the bare quark masses. We used the
R algorithm of the MILC collaboration's code \cite{milc}.

\TABLE{\begin{tabular}{l|l|l|l|l}
$Re(\mu)$& 0.1 & 0.2 & 0.3 & 0.4 \\
\hline
Re($\beta_0$); $L_s=4$ &
5.151(1) & 5.141(1) & 5.127(2) & 5.121(5)  \\
$10^2$Im($\beta_0$) &
5.56(8) & 5.50(9) & 5.42(15) & 5.56(38)  \\
\hline
Re($\beta_0$); $L_s=6$ &
5.193(1) & 5.174(1) & 5.152(3) & 5.143(7)  \\
$10^2$Im($\beta_0$) &
2.66(6) & 2.54(9) & 2.19(31) & 1.82(39)  \\
\hline
Re($\beta_0$); $L_s=8$ &
5.193(1) & 5.172(1) & 5.159(1) & 5.140(1)  \\
$10^2$Im($\beta_0$) &
1.38(6) & 1.32(17) & 1.31(18) & 0.48(7)  \\
\hline
Re($\beta_0$);$L_s\rightarrow \infty$ &
5.201(1) & 5.178(1) & 5.162(2) & 5.143(2)  \\
$10^2$Im($\beta_0$) &
1.02(6) & 1.12(11) & 0.74(19) & -0.25(10)  \\
\hline\hline
$\beta$ & $m_\pi$ & $m_\rho$ & $R_0$ & $\sqrt{\sigma}$ \\
\hline
5.208   & 0.393(2)       & 1.22(2)        & 1.87(3)     & 0.58(7)       \\
5.164   & 0.393(2)       & 1.28(3)        & 1.76(5)     & 0.75(5)      \\
\end{tabular}
\vspace{0.3cm}
\caption{\label{zeros}
$T\neq 0$ and $T=0$ results. The upper part is a
Summary of the Lee-Yang zeros obtained at different chemical
potentials, 
%(4,6,8 and $\infty$ indicate the spatial extensions --and their
%extrapolation-- of our $L_t=4$ lattices), 
while the lower part shows the
measured $T=0$ observables for two $\beta$ values.
}}

\FIGURE{\epsfig{file=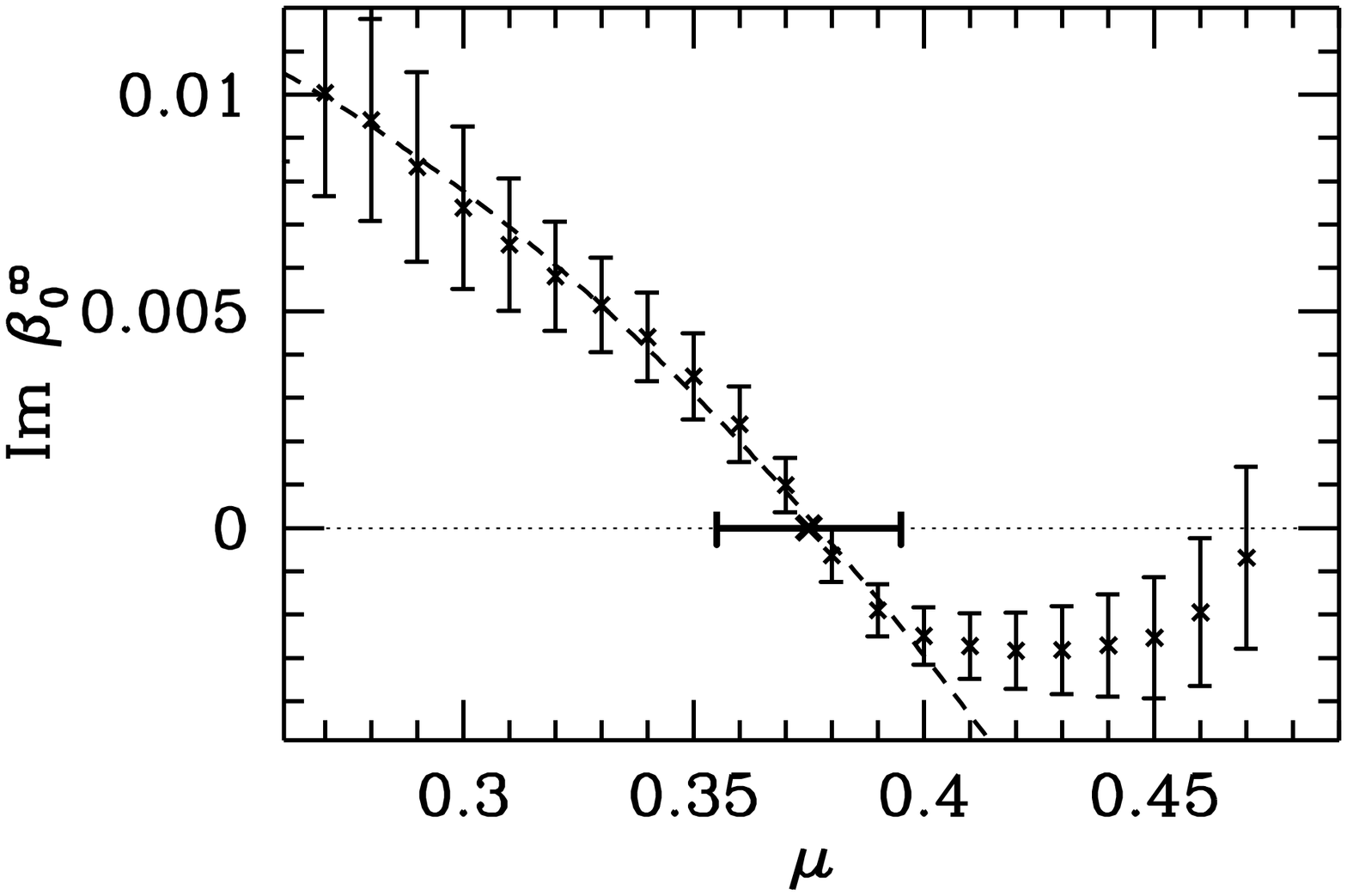,width=7.7cm}
\caption{\label{infV}
Im($\beta_0^\infty$) 
as a function of the chemical potential. 
}}

At  $T\neq 0$
we determined the complex valued Lee-Yang zeros, 
$\beta_0$, for different V-s 
as a function of $\mu$. Their 
V$\rightarrow \infty$ limit was given by a 
$\beta_0(V)=\beta_0^\infty+\zeta/V$
extrapolation. The results (listed in Table \ref{zeros}) 
are from 14000, 3600 and 840 configurations on 
$L_s$=4,6 and $8$ lattices, respectively. Figure \ref{infV}
shows Im($\beta_0^\infty$) as a function of $\mu$ enlarged around 
the endpoint $\mu_{end}$.
The picture
is simple and reflects the physical expectations. For small
$\mu$-s the extrapolated Im($\beta_0^\infty$) is inconsistent with
a vanishing value, and the prediction is a crossover.
Increasing $\mu$ the value of Im($\beta_0^\infty$) decreases, 
thus the transition becomes consistent with a first order phase
transition. (Note, that errors decrease close to the endpoint,
and the Im($\beta_0^\infty$) extrapolation, due to the relatively small 
volumes, slightly overshoots. 
Both phenomena were observed already in the electroweak case e.g.
\cite{4ewpt}). 
The statistical error
was determined by a jackknife analysis using subsamples of the
total $L_s=4,6$ and $8$ partition functions. The 
systematic uncertainty, estimated from the overshooting, was added 
linearly to the statistical error. 
The dashed line of Figure \ref{infV} shows the fit and  
leads to our primary result: $\mu_{end}=0.375(20)$. 
      
Table \ref{zeros} contains also the 
$T=0$ results. To set the physical scale we used a 
weighted average of $R_0$ (1/403~MeV), $m_\rho$ (770~MeV) and 
$\sqrt{\sigma}$ (440~MeV), obtained in lattice units
by different fitting procedures.
It is important to note, that (including systematic errors due to 
finite V) we have 
$(R_0\cdot m_\pi)=0.73(6)$, which is at least twice
as large as its physical value. Thus our $m_{u,d}$ is
at least four times larger than it should be.

\FIGURE{\epsfig{file=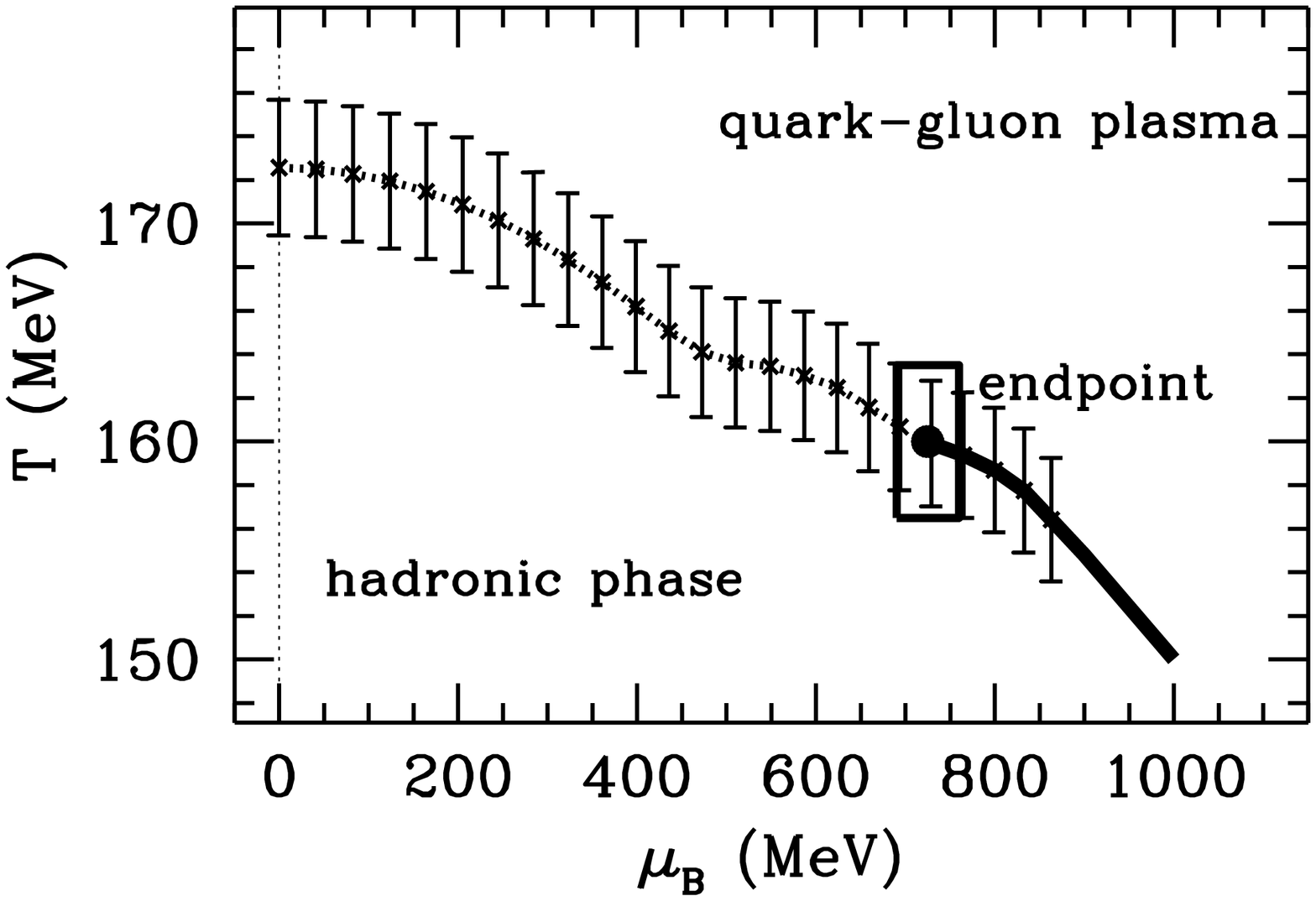,width=7.7cm}
\caption{\label{physical}
The phase diagram in physical units. Direct results are with errorbars.
Dotted line illustrates the crossover, solid line the first order phase
transition. The small box shows the uncertainties of the endpoint.
}}

Let us estimate the applicability of the method approaching the
chiral and continuum limits.
In the present analysis the evaluation of the eigenvalues was
somewhat less costly than the production of the configurations.
For physical $m_{u,d}$ the latter would need an additional 
factor of ${\cal O}(50)$ (the former remains the same).
Thus, for physical masses at least upto V=4$\cdot 12^3$  
the cost of the eigenvalue determination is subdominant. Extending the
analysis to this volume reduces the error on $\mu_{end}$ 
to a level, which is not even needed (uncertainties due to
finite lattice spacing could be more important).
Since for finer lattices the eigenvalue evaluation 
goes with $L_s^9$ and the 
configuration production at least with $L_s^9$ the
eigenvalue evaluation remains subdominant. 
At physical masses $\mu_{end}$ is probably closer 
to the $\mu$=0 axis (for recent lattice works see \cite{B90}). 
Thus, the overlap between $\mu$=0 and $\mu \neq 0$ configurations
is even better. It means less statistic might be enough
to apply eq. (\ref{reweight}) than it was used in this work.  
Note, that the quark masses of the present work are half of
those used in Ref. \cite{FK01}; however, in both cases
it was possible to reweight in $\mu$ far beyond $m_\pi/2$ (the
typical premature onset $\mu$ value of the Glasgow method 
\cite{glasgow}). Thus, we expect that our method
does not suffer from this type of onset problem when approaching
the chiral limit.

Figure \ref{physical} shows the phase diagram in
physical units, thus
$T$ as a function of $\mu_B$, the baryonic chemical potential 
(which is three times larger then the quark chemical potential). 
The endpoint
is at $T_E=160 \pm 3.5$~MeV, $\mu_E=725 \pm 35$~MeV.
At $\mu_B$=0 we got $T_c=172 \pm 3$~MeV.

\section{Conclusions, outlook.}

We used a recently proposed method \cite{FK01} 
and studied the $\mu$-$T$ phase diagram of QCD with 
dynamical $n_f$=2+1 quarks. We presented an {\it ab initio}
technique to determine the 
location of the endpoint. Using the above method we obtained 
$T_E$=160$\pm$3.5~MeV for the temperature
and $\mu_E$=725$\pm$35~MeV for the baryonic chemical potential
of the endpoint. 
This result was based on the 
V$\rightarrow \infty$ behavior of the Lee-Yang zeros of the 
partition function. We used $L_t$=4 and 
our quark masses were ``semi-realistic'' ($m_s$ 
was set to about its physical value, whereas 
$m_{u,d}$ were four times heavier
than in the real world). Though $\mu_E$ is too
large to be studied at RHIC/LHC, the endpoint would 
probably move closer to the $\mu$=0 axis 
when the quark masses get reduced. 
At $\mu$=0 we obtained $T_c$=172$\pm$3~MeV.
Clearly, more work is needed to get
the final values. One has to  extrapolate to zero step-size
in the R-algorithm and to the thermodynamic, 
chiral and continuum limits.    

We thank F. Csikor, F. Karsch, I. Montvay and A. Ukawa for useful  
comments on the manuscript. 
This work was partially supported by Hungarian Science Foundation
grants No. 
OTKA-34980/\-29803/\-22929/\-28413/\-OMMU-708/\-IKTA. 
This work was in part based 
on the MILC collaboration's public lattice gauge theory code \cite{milc}.

\end{document}